# A comparative analysis of plasmonic and dielectric metasurface sensing platforms powered by bound states in the continuum

*Tao Jiang, Angana Bhattacharya, Martin Barkey, Andreas Aigner, Thomas Weber, Juan Wang, Stefan A. Maier, and Andreas Tittl\**

T. Jiang, A. Bhattacharya, M. Barkey, A. Aigner, T. Weber, J. Wang, A. Tittl
Chair in Hybrid Nanosystems, Nano-Institute Munich, Faculty of Physics, Ludwig-Maximilians-University Munich, Munich, 80539, Germany.
E-mail: andreas.tittl@physik.uni-muenchen.de

S A. Maier
School of Physics and Astronomy, Monash University, Clayton, Victoria 3800, Australia
Department of Physics, Imperial College London, London SW7 2AZ, United Kingdom

**Funding**: This project was funded by the Deutsche Forschungsgemeinschaft (DFG, German Research Foundation) under grant numbers EXC 2089/1–390776260 (Germany's Excellence Strategy) and TI 1063/1 (Emmy Noether Program), the Bavarian program Solar Energies Go Hybrid (SolTech), and Enabling Quantum Communication and Imaging Applications (EQAP), and the Center for NanoScience (CeNS). Funded by the European Union (ERC, METANEXT, 101078018 and EIC, NEHO, 101046329). Views and opinions expressed are however those of the author(s) only and do not necessarily reflect those of the European Union, the European Research Council Executive Agency, or the European Innovation Council and SMEs Executive Agency (EISMEA). Neither the European Union nor the granting authority can be held responsible for them. S.A.M. additionally acknowledges the Lee-Lucas Chair in Physics.

**Abstract**

Nanophotonic platforms based on surface-enhanced infrared absorbance spectroscopy (SEIRAS) have emerged as an effective tool for molecular detection. Sensitive nanophotonic sensors with robust resonant modes and amplified electromagnetic near fields are essential for spectroscopy, especially in lossy environments. Metasurfaces driven by bound state in the continuum (BICs) have unlocked a powerful platform for molecular detection due to their exceptional spectral selectivity. While plasmonic BIC metasurfaces are preferred for molecular spectroscopy due to their high surface fields, enhancing the interaction with analytes, dielectric BICs have become popular due to their high-quality factors and, thus high sensitivity. However, their sensing performance has largely been demonstrated in air, neglecting the intrinsic infrared (IR) losses found in common solvents.

This study evaluates the suitability of plasmonic versus dielectric platforms for in-situ molecular spectroscopy. Here, the sensing performance of plasmonic (gold) and dielectric (silicon) metasurfaces is assessed across liquid environments with varying losses resembling typical solvents. The results show that dielectric metasurfaces excel in dry conditions, while plasmonic BIC metasurfaces outperform them in lossy solvents, with a distinct crossover point where both show similar performance. Our results provide a framework for selecting the optimal metasurface material platform for SEIRAS studies based on environmental conditions.

*Tao Jiang and Angana Bhattacharya contributed equally to this work.*

## 1. Introduction

Developing compact and highly sensitive sensors is crucial for advancing next-generation biosensing techniques across various fields, including clinical diagnostics, virus detection, drug screening, and cellular secretion analysis, as well as for the realization of rapid and reliable point-of-care (POC) testing.[1–6] Nanophotonic sensors offer several advantages for molecular sensing including label-free, non-invasive detection,[2] surface-enhanced spectroscopy to identify molecular signatures,[7] and on-chip sensing capabilities through miniaturized designs.[8] Vibrational spectroscopic techniques, such as surface-enhanced infrared absorbance spectroscopy (SEIRAS) and surface-enhanced Raman spectroscopy (SERS) have enabled sensitive detection and molecular fingerprinting of trace analytes by enhancing light-matter interactions through coupling molecular vibrations to photonic resonators.[9–11] In the mid-infrared (mid-IR) region, SEIRAS has proven effective for detecting chemically specific absorption bands of minute quantities of analytes, proteins, lipids, and other biomolecules.[10,12–14] A variety of nanophotonic geometries have been explored for SEIRAS, including nanorods,[15] nano-gap structures,[8,16–18] metal-insulator-metal (MIM) designs,[5,19,20] and plasmonic metasurfaces.[21–27] Gold-based plasmonic resonators have long been the standard for nanophotonic sensing due to their strong surface fields, leading to effective molecular coupling and excellent surface selectivity. However, plasmonic nanostructures suffer from high intrinsic ohmic losses which limits their performance.[28–32] As an alternative, high-refractive-index dielectric materials have been investigated for molecular sensing, where low intrinsic losses enabled high-quality (Q) factor resonances that are highly responsive to environmental changes.[33–35] Studies have shown that dielectric metasurfaces outperform plasmonic ones in air.[36–38] However, biomedical assays, clinical diagnostics, and cell secretion monitoring often occur in solvents, and measurements in air overlook the infrared (IR) absorption losses inherent in such solvents. Hence, it remains unclear whether plasmonic or dielectric platforms perform better in such environments. Therefore, a direct comparison of their sensing performance in lossy solvents is critical for determining the optimal material platform for practical sensing applications. Despite its importance, only a few studies have reported molecular sensing in lossy, solvent-based environments.[24,32] A key challenge is understanding how the sensitivity of the mode relates to its robustness when exposed to lossy environments. Important parameters in such a comparison include resonance frequency, intrinsic and radiative losses, and molecular coupling. To enable a fair comparison, a platform is needed where the resonance frequency and radiative losses of metasurfaces composed of different materials can be independently controlled. Bound states in the continuum (BIC) metasurfaces serve as an effective tool here.

BIC metasurfaces have been successfully used in SEIRAS-based sensors providing resonances with high Q-factors and excellent spectral selectivity.[35,39–41] While true BIC modes are non-radiative and do not couple to the far field, quasi-BIC (qBIC) modes allow control of the radiative losses by adjusting the asymmetry of the resonators.[31] By employing BIC metasurfaces to compare the sensing performance between dielectric and plasmonic platforms, the radiative loss and resonance frequency can be tuned individually, allowing a comparison primarily driven by material performance.

Here, we compare the sensing performance of plasmonic (gold) and dielectric (silicon) BIC metasurfaces in realistic, lossy solvents in the mid-IR range. Through comprehensive numerical simulations and experimental measurements, we examine the molecular absorbance signal of an exemplary analyte layer (PMMA) in different environments, including heavy water ($D_2O$), water ($H_2O$), and dimethyl sulfoxide (DMSO), each representing varying degrees of optical loss. To ensure precise and controlled solvent delivery, experiments are performed using a microfluidic setup. By optimizing the radiative loss through tailored asymmetries for both the plasmonic and dielectric metasurfaces in each sensing environment, the effects of radiative losses could be excluded from the study. Our findings reveal that there is no universally superior material platform for all environments, and the sensitivity of the material platform depends strongly on the sensing environment. While it was observed that dielectric metasurfaces are more sensitive in loss-less media such as air, plasmonic metasurfaces demonstrate higher sensitivity in moderately lossy solvents ($k_{env} > 2 \times 10^{-3}$). Moreover, there exists a clear crossing over point where the sensitivity of the material platform changes from dielectric to plasmonic depending on the solvent losses. For highly lossy solvents, both material platforms demonstrate similar performance indicating minimal dependence of the material platform on the sensing environment. Experimentally, this trend is confirmed for solvents such as $D_2O$ and DMSO, both widely used in practical biosensing applications.

## 2. Results and Discussion

### 2.1. Metasurface design and sensing setup

We investigate sensing processes in media with varying losses. In addition to damping the resonance, loss variations also cause spectral shifts of the resonances, detuning the resonances from the spectral position of the target absorption bands of the analyte. This can be addressed by using multispectral metasurfaces with resonances covering a range of wavelengths around the absorption band, for example using pixelated geometries.[32,37] However, achieving this with a conventional pixelated metasurface would pose significant challenges, requiring highly precise tuning of pixel parameters to maintain a constant resonance frequency for each incremental environmental variation. The low resonance density of pixelated metasurfaces also adds to reduced spectral resolution, making it difficult to detect the target absorption band of the analyte.

To address this, we employed a gradient BIC metasurface.[42,43] Instead of arranging elliptical resonators in a 2D periodic array, we introduced a multiplicative lateral scaling factor 'S' which continuously modifies the unit cell dimensions to produce a spectral resonance tuning from 1300 cm$^{-1}$ to 1800 cm$^{-1}$, ensuring that both the gold and silicon BIC metasurfaces consistently achieve resonances at the desired excitation frequency despite changes in the surrounding environment.

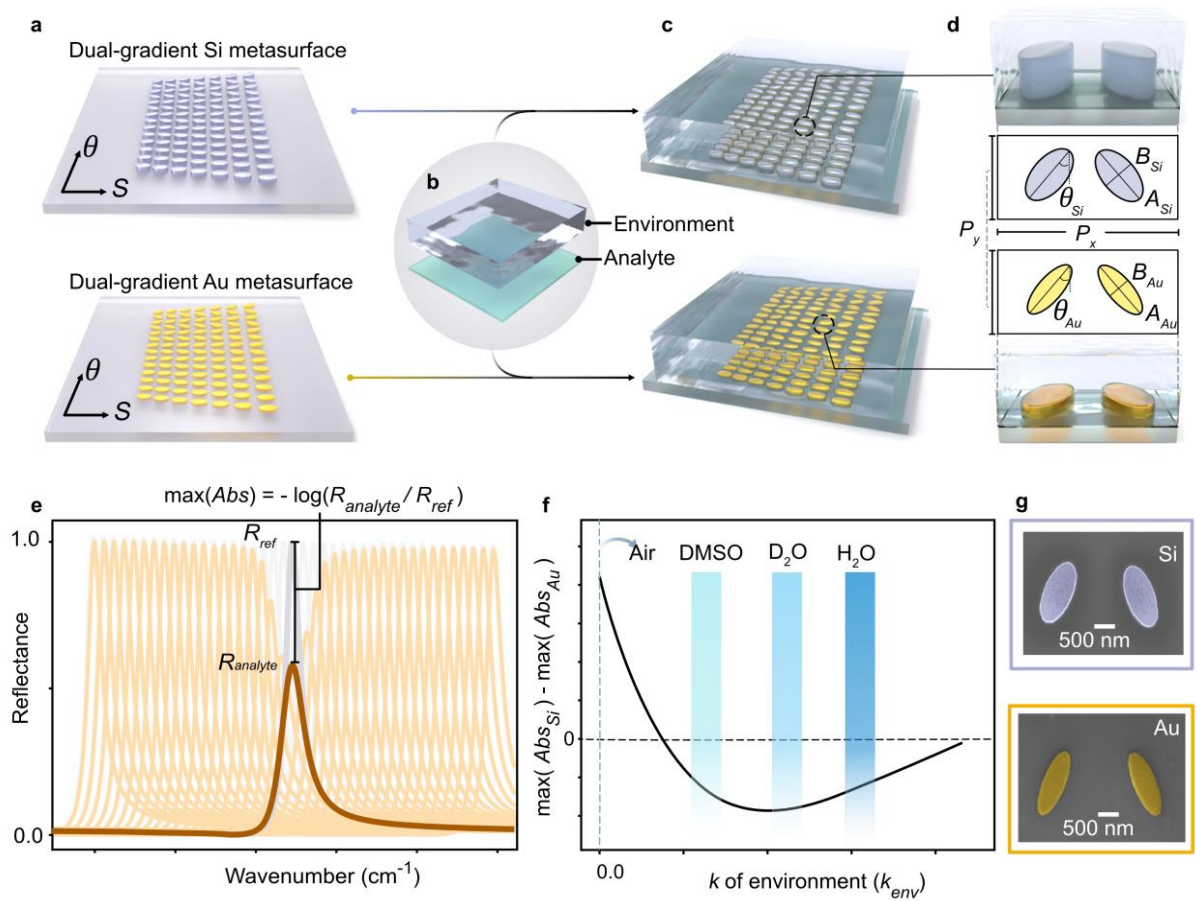

**Figure 1. Comparison of Si and Au metasurfaces for sensing performance**. (a) Schematic of the silicon (Si) and gold (Au) metasurfaces, incorporating a spectral gradient (scaling factor, $S$) along the horizontal axis and a coupling gradient (tilting angle, $\theta$) along the vertical axis. (b)

Schematic of the metasurfaces coated with analyte (PMMA) and immersed in an environment with a refractive index $n + i\, k_{env}$. (c) Illustration of the sensing setup. (d) Unit cell geometry: the Si and Au metasurfaces share the same lattice periods $(P_x, P_y)$. $A_{Si}$ ($A_{Au}$) and $B_{Si}$ ($B_{Au}$) denote the major and minor axes of the elliptical resonators, while $\theta_{Si}$ ($\theta_{Au}$) represents the tilt angle. The Si ellipses have a height of $H_{Si}$ = 750 nm, whereas the Au ellipses have height $H_{Au}$ = 100 nm. (e) Reference spectra, $R_{Ref}$ (gray), of the bare metasurface and reflectance spectra of the analyte-coated metasurface, $R_{analyte}$ (orange), immersed in the lossy environment. Absorbance (*Abs*), defined as $Abs = -\log(R_{analyte}/R_{Ref})$, is considered as the sensing parameter. (f) Difference in maximum absorbance between Si and Au metasurfaces, max($Abs_{Si}$) - max($Abs_{Au}$), as a function of the environmental loss ($k_{env}$) of the surrounding medium. (g) Scanning electron microscope (SEM) images of the fabricated Si and Au metasurfaces.

**Figure 1a** shows the Si and Au dual-gradient metasurfaces, with S and tilting angle '$\theta$' varying along perpendicular directions. The tilting angle breaks the symmetry of the BIC mode into a quasi-BIC (qBIC) mode.[37] After coating with an analyte layer the metasurface is immersed in a lossy environment to evaluate its sensing response. Polymethyl methacrylate (PMMA) was chosen as the analyte because it interacts similarly with both gold and silicon (Figure 1b). The geometric parameters were optimized to achieve a resonance at 1730 cm$^{-1}$ the absorption peak of PMMA. The gradient metasurfaces are immersed in the sensing environment, as depicted in Figure 1(c). We chose a tilted-ellipse-based BIC-metasurface design because of it stability against fabrication-induced geometrical variations and low baseline reflectance, making it effective for molecular detection in the mid-IR range (1300 cm$^{-1}$ to 1800 cm$^{-1}$).[35,44] The asymmetry is dictated by $\theta$ (Figure 1d), where $\theta = 0°$ corresponds to the true BIC condition. Breaking the in-plane symmetry excites quasi-BIC (qBIC) resonances with sharp and spectrally selective resonances.[45] We selected infrared-transparent calcium fluoride (CaF$_2$) as the substrate and fabricated two metasurfaces: one composed of gold (Au) resonators and the other of silicon (Si). To ensure a fair comparison, both metasurfaces were designed with nearly identical structural parameters, maintaining a pitch of $P_x$ = 4000 μm and $P_y$ = 2400 μm for both platforms, and similar long (A), and short diameters (B). These parameters are depicted in Figure 1(d). The primary difference is the height of the nanostructures, with 100 nm for the Au metasurface and 750 nm for the Si metasurface. We denote the reflectance of the metasurface without an analyte layer as $R_{ref}$, while the reflectance after the deposition of an ultrathin (5 nm) analyte layer is represented as $R_{analyte}$. The sensing performance is quantified by the logarithmic ratio between $R_{analyte}$ and $R_{ref}$ defined as absorbance, $Abs = -\log(R_{analyte}/R_{ref})$, as illustrated in Figure 1(e). We analyze the response of the metasurface under varying refractive indices of the external environment, represented as $n + i\, k_{env}$. For ease of numerical analysis, the real part of the refractive index ($n$) is held fixed at 1.33. The imaginary component ($k_{env}$), which accounts for medium-induced losses, is systematically varied from 0 (lossless, e.g., air) to 0.1 (highly lossy). Figure 1(g) presents exemplary scanning electron microscope (SEM) images of the fabricated Si and Au metasurfaces (for fabrication details see Methods).

## 2.2. Numerical comparison of bare metasurfaces in lossy media

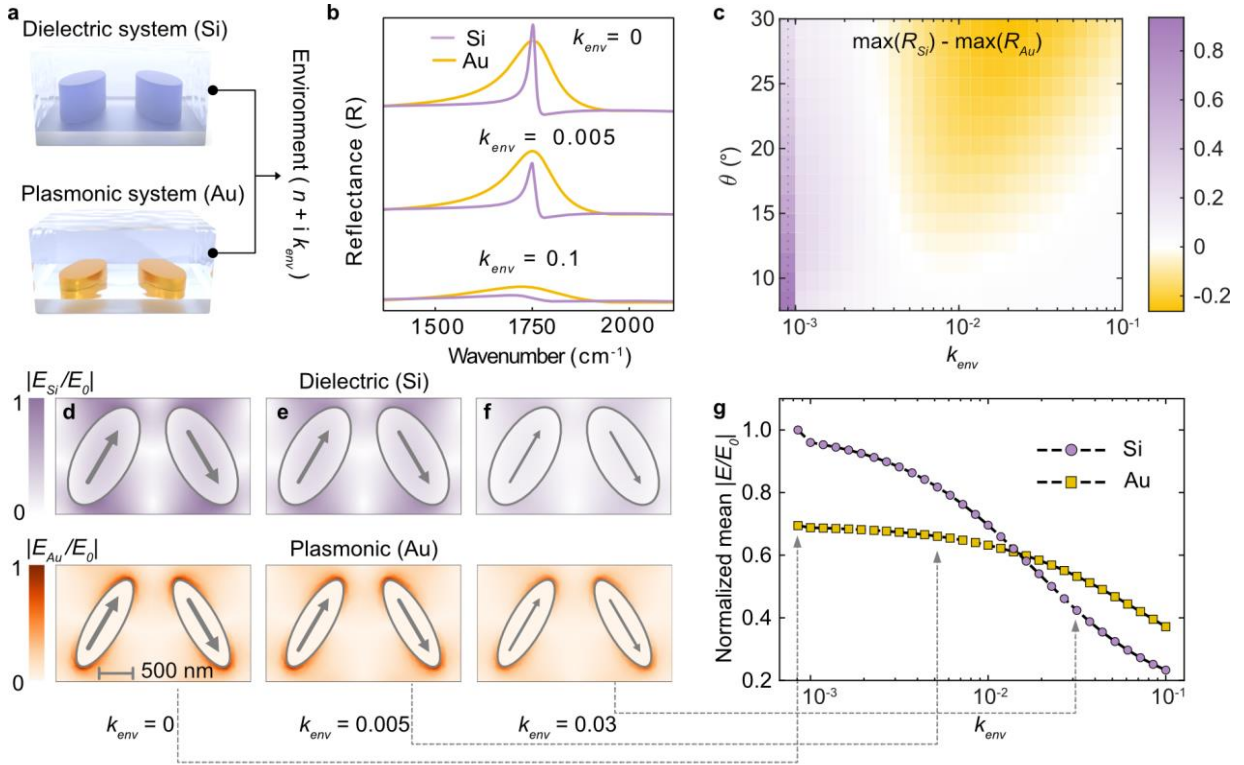

**Figure 2. Metasurface response to environmental losses without analyte layer**. (a) Schematic of the plasmonic (Au) metasurface and dielectric metasurface (Si) in a lossy environment of refractive index $n + i\, k_{env}$. (b) Numerically simulated reflectance of the Si (purple) and Au (yellow) metasurfaces for a constant $n = 1.33$ and $k_{env} = 0$ (corresponds to a loss-less environment), $k_{env} = 0.005$ (corresponds closely to the value of heavy water (D$_2$O)), and $k_{env} = 0.1$ (corresponds to a highly lossy environment). (c) Difference in the peak reflectance (max($R$)) between Si and Au metasurface as a function of changing $\theta$ and $k_{env}$. Dashed line indicates the $k_{env}$ at zero value. Electric field enhancement for the Si and Au metasurfaces for $\theta = 30°$ at (d) at $k_{env} = 0$, (e) $k_{env} = 0.005$, (f) $k_{env} = 0.03$. (g) Normalized mean electric field enhancement ($|E/E_0|$) of the Au (yellow) and Si (purple) metasurfaces.

We first numerically investigated the resonance behavior of the bare metasurfaces in the imaginary medium composed of changing $k_{env}$ (**Figure 2a**). Figure 2b shows the reflectance spectra for the Au metasurface (yellow) and Si metasurface (purple) at representative values of $k_{env} = 0, 0.005, 0.1$. Notably, $k_{env} = 0.005$ closely corresponds to the experimentally determined mid-IR value of $k_{env}$ for $D_2O$. For a low loss environment ($k_{env} = 0$), the Si metasurface exhibits a sharper resonance linewidth and a higher reflectance peak as compared to the Au metasurface. However, as $k_{env}$ increases to 0.005, the maximum reflectance, (max($R$)), which indicates the peak value of the reflectance spectrum, decreases for Si while that for Au surpasses that of Si. At $k_{env} = 0.1$, the reflectance peaks of both material platforms are significantly suppressed by the strong environmental loss. To gain further insight into the interplay between environmental loss and asymmetry, we computed max($R$) for each metasurface across asymmetries ($\theta$) for values ranging from 5° to 30°. The difference in maximum reflectance, max($R_{Si}$) – max($R_{Au}$), is color-mapped in Figure 2c where purple color signifies max($R_{Si}$) > max($R_{Au}$), yellow indicates max($R_{Si}$) < max($R_{Au}$), and white denotes equal max($R$) for both Si and Au metasurfaces. The asymmetry introduced by the tilting angle $\theta$ represents the radiative loss in the BIC metasurface.[31] A fine balance between radiative loss and reflectance amplitude is essential for an efficient study. Increasing $\theta$ raises reflectance amplitude, whereas increasing environmental losses reduces it. At small tilt angles, corresponding to low radiative losses, sharp resonance peaks with low reflection amplitudes are observed. For $\theta < 10°$, the Si metasurfaces show better performance in low-loss environment, particularly when $k_{env} = 0$. However, both metasurfaces have comparable performance due to the induced loss in the environments, as indicated by the white region in the colormap. At $k_{env} = 0.005$ and, $\theta = 10^0$, the maximum reflectance of Si (light purple line) and Au (light yellow line) overlap, indicating comparable performance. By contrast, as values of $\theta$ increases to 20°, Au outperforms Si for lossy environments and at high asymmetries ($\theta = 30°$), the difference for max($R$) grows even more pronounced. Furthermore, we simulated the electric fields at the resonance wavenumber of 1730 cm$^{-1}$ for $k_{env} = 0$ (Figure 2d), 0.005 (Figure 2e), and 0.3 (Figure 2f). For $k_{env} = 0$, the electric field for the Si resonator is spread over the entire ellipses, while localized hotspots of electric fields at the tips of the ellipses are observed for the Au resonators. As $k_{env}$ is increased, the overall field concentration for the Si resonators reduces considerably, being completely quenched for $k_{env} = 0.3$. On the other hand, the field hotspots for the Au resonators remain localized despite increasing environmental losses, and the presence of hotspots are observed even for high environmental loss ($k_{env} = 0.3$). The normalized mean electric field enhancement for both Au and Si metasurfaces is plotted against increasing $k_{env}$, as shown by the yellow (Au) and purple (Si) lines in Figure 2g. A crossover point occurs for $k_{env}$ > $1.2 \times 10^{-2}$, beyond which the mean surface field of Au exceeds that of Si. Thus, a steep leakage of the mean fields is observed for the Si metasurfaces with increasing $k_{env}$, as opposed to the Au metasurfaces where the decrease in the mean electric field enhancement with $k_{env}$ is relatively stable.

## 2.3. Numerical comparison of sensing performance for analyte coated metasurfaces

Next, we look into the sensing performance of the metasurfaces. We numerically simulated a thin analyte layer (PMMA) covering the Si and Au metasurfaces uniformly and then immerse the metasurfaces in the solvent. As introduced previously (Figure 1e), the sensing performance is quantified by the absorbance, defined as the logarithmic ratio $Abs = -\log R_{analyte}/R_{Ref}$. This metric captures the amplitude difference before and after analyte binding and thus enables precise tracking of molecular infrared absorbance spectra. The peak value max($Abs$) represents the highest variation in reflectance caused by the analyte and thus represents the point of highest sensitivity of the metasurface. To systematically assess the sensing performance, we computed max($Abs$) at the PMMA absorption peak (1730 cm$^{-1}$) across a range of asymmetries ($\theta$) and varying $k_{env}$ from 0 to 0.1. We further computed the difference in the maximum absorbance between Si and Au metasurfaces, given by max($Abs_{Si}$) – max($Abs_{Au}$). This is mapped in **Figure 3**a, where purple color corresponds to max($Abs_{Si}$) > max($Abs_{Au}$) indicating better sensitivity for Si, yellow corresponds to max($Abs_{Si}$) < max($Abs_{Au}$) indicating higher sensitivity for Au, while white indicates similar sensing performance in both material platforms. We observe that at very low environmental losses ($k_{env}$ < 10$^{-3}$), Si exhibits higher sensitivity across the full range of $\theta$. However, as $k_{env}$ increases, there is a clear transition, where Au gradually outperforms Si. For high loss, i.e., $k_{env}$ = 0.1, both metasurfaces show similar sensing performance with resonance amplitudes strongly quenched, as shown by the light yellow regions in Figure 3a. Even though this 2D map provides a general idea of the sensing performance of the systems, it does not take into account the optimal metasurface configuration. By carefully tuning the tilt angle $\theta$, we can control the radiative losses in the system, allowing us to choose the asymmetry that maximizes sensitivity under each loss condition. In doing so, the influence of radiative loss is kept consistent for both Si and Au, and we can focus primarily on the effect of the intrinsic losses of both platforms.

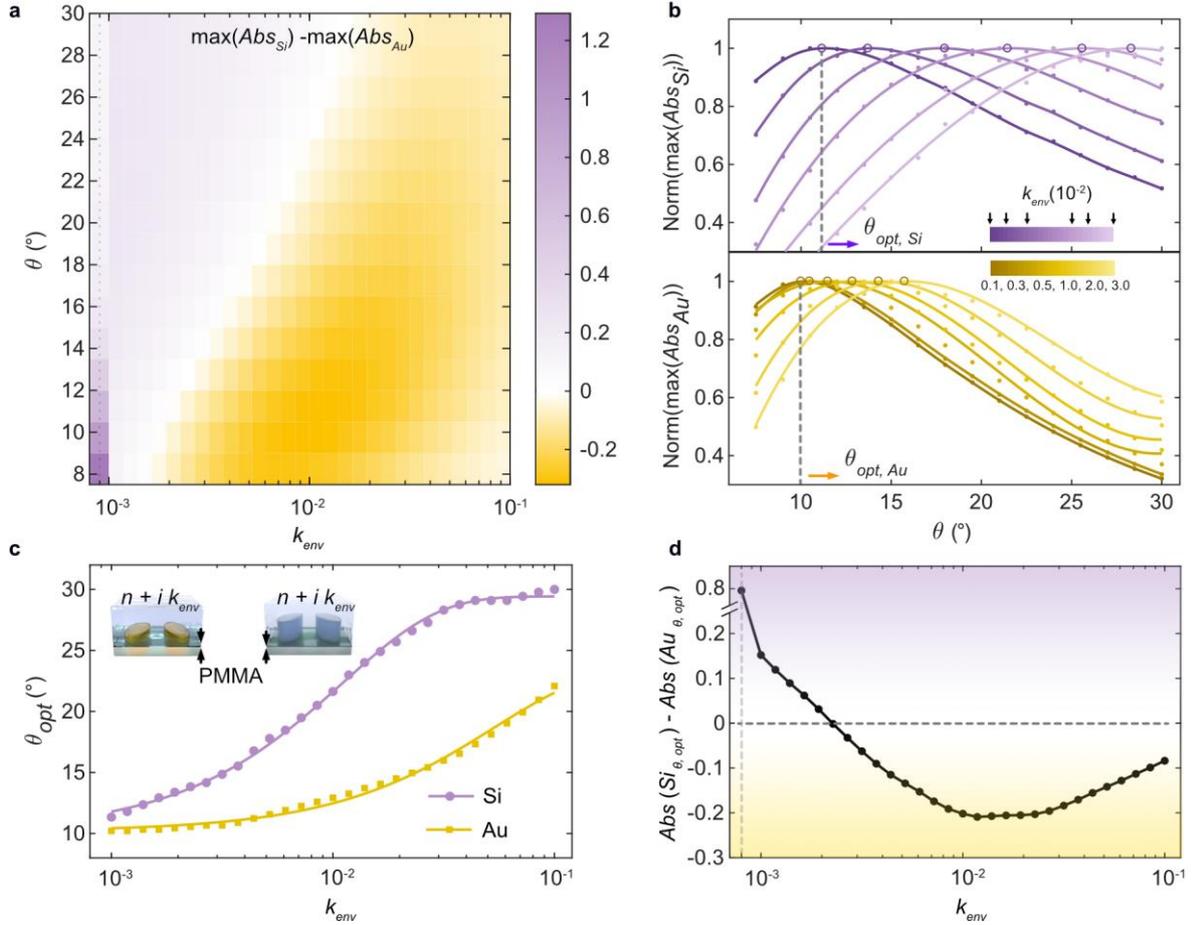

**Figure 3**: **Metasurface sensing of analyte layer in lossy environments**. (a) Difference in the maximum value of the sensing metric, *Abs*, between Si and Au as a function of changing $\theta$ (°) and $k_{env}$. Dashed line indicates the $k_{env}$ at zero value. (b) Normalized spectra of max(*Abs*) with changing $\theta$ (°) for Si (purple) and Au (yellow) metasurfaces. The value of $\theta$ corresponding to the peak of the normalized max(*Abs*) at a fixed valued of $k_{env}$ represents the optimal asymmetry, $\theta_{opt}$, for maximized sensitivity in that environment. (c) Plot of $\theta_{opt}$ (°) vs $k_{env}$. (d) Difference in *Abs* calculated at $\theta_{opt}$ between Si and Au metasurfaces, represented as $Abs(Si_{\theta,opt}) - Abs(Au_{\theta,opt})$ as a function of $k_{env}$.

Figure 3b presents a magnified view of the normalized max(*Abs*) as a function of $\theta$, highlighting the variation with $k_{env}$ in the range 0.1 to 3.0×10⁻², to facilitate clearer interpretation of the angular dependence. The optimally sensitive asymmetry angle, $\theta_{opt}$ (°), is defined as the value of $\theta$ corresponding to the peak of the normalized max(*Abs*) curve at a given $k_{env}$. The dashed lines in figure 3b indicate the extracted $\theta_{opt}$ values corresponding to $k_{env} = 0$ for Si and Au metasurfaces. The extracted values of $\theta_{opt}$ for each $k_{env}$ are presented in figure 3c, enabling identification of the most sensitive asymmetry configuration for both Si and Au metasurfaces at a given $k_{env}$. Notably, the plot reveals that $\theta_{opt}$ differs for a given $k_{env}$ between the two material platforms, indicating that a fixed asymmetry does not yield the same sensitivity across Si and Au metasurfaces. For example, from figure 3c, we observe that at $k_{env} = 0.005$, the optimal tilting angle $\theta_{opt}$ for Si metasurface is 17.2° while that for Au metasurface it is 11.4°. To compare the best possible sensing performance for each material, we thus calculate max($Abs$) at these individual $\theta_{opt}$ for the Si and Au metasurfaces and plot the resulting difference, $Abs(Si_{\theta opt})$ - $Abs(Au_{\theta opt})$ in Figure 3d, where the values of $k_{env}$ varies from 0 to 0.1. It is observed that for $k_{env} < 2.0 \times 10^{-3}$, Si has superior sensing performance than Au. At $k_{env} \approx 2.0 \times 10^{-3}$, a transition point is reached where Si and Au performs similarly, and as $k_{env}$ increases beyond $2.0 \times 10^{-3}$, Au outperforms Si, depicted by the negative values in the optimized $Abs$ difference. This range of $k_{env}$ closely matches experimentally relevant solvents used for dissolving biomolecules, including $D_2O$ and water. As $k_{env}$ is increased to very high values ($k_{env} = 0.1$), the difference in $Abs$ tends to 0. Hence, this implies that in highly lossy environments, the choice of material platform becomes considerably less relevant.

## 2.3. Experimental comparison of sensing performance

To validate our numerical findings, we conducted an experimental study to analyze the performance of fabricated metasurfaces for analyte sensing in different solvent environments **(Figure 4)**. By incorporating a scaling factor 'S' along the x-direction (as discussed in section 2.1), we enabled accurate measurement of the resonance for each solvent, overcoming shifts due to the refractive index of the solvent and analyte (**Figure S1**). Additionally, the variation in $\theta$ along the y-direction of the metasurfaces allows access to the optimal absorbance condition across the metasurface. Each metasurface was placed into a microfluidic cell and reflectance measurements were performed Measurements were initially conducted in air for both the Si and Au metasurfaces. Next, a set of solvents, including DMSO, $D_2O$, and $H_2O$ were sequentially introduced into the microfluidic cell, and reflectance measurements were recorded for each solvent, yielding normalized reference spectra $R_{ref}$. Next, each metasurface was coated with a thin layer of PMMA as the analyte (for details see Methods and **Figures S2, S3**) and the coated metasurface was remounted in the microfluidic cell.

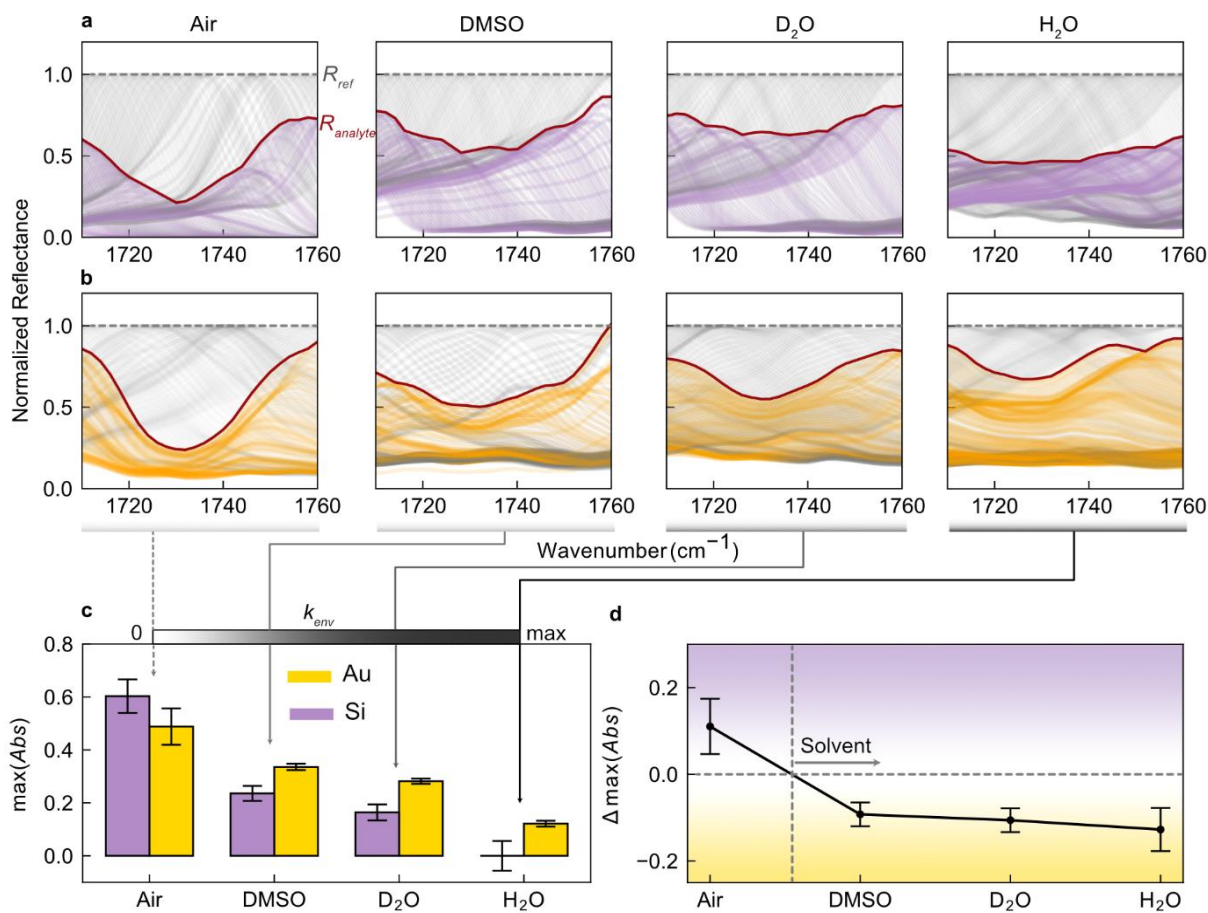

**Figure 4**: **Experimental evaluation of metasurface sensing performance at 1730 cm$^{-1}$ for different solvents**. Normalized reflectance spectra for the (a) Si and (b) Au metasurfaces measured in air, DMSO, D$_2$O and H$_2$O. The dashed line shows the envelope of the normalized reference reflectance ($R_{Ref}$); the red solid line shows the envelope of the normalized reflectance with an analyte ($R_{analyte}$). (c). Mean absorbance values at 1730 cm$^{-1}$ with standard deviation error bars calculated from the 50 highest measured *Abs* values for each solvent. (d) Difference in mean absorbance from the 50 highest *Abs* values between Si and Au metasurfaces.

The solvents were sequentially flowed over the coated metasurface, with the resulting normalized reflectance spectra referred to as $R_{analyte}$. Figure 4(a,b) shows the normalized reflectance spectra before and after PMMA coating, respectively (for unprocessed spectra refer to **Figure S4**). The reflectance spectra are normalized to the peak reflectance of the uncoated metasurface for each solvent. A distinct dip appears at 1730 cm$^{-1}$ in the envelope of $R_{analyte}$, corresponding to the characteristic vibrational absorption of PMMA. As the $k_{env}$ increases, the modulation depth of the amplitude of the $R_{analyte}$ decreases for both of metasurfaces. This reduction in modulation is attributed to increased optical losses in the solvent environment, which suppress the qBIC resonance and thus reduce the sensing capability. Notably, the characteristic analyte absorption feature for the Si metasurface in $H_2O$ becomes indistinct at 1730 cm$^{-1}$. This indicates a strong suppression of the qBIC resonance, resulting in an absorbance value close to zero for the Si metasurface in water, as seen in Figure 4(b). For an accurate analysis, the 50 highest measured *Abs* values corresponding to each solvent were chosen and the average of these selected values are plotted in Figure 4(c). With increasing $k_{env}$, the absorbance of the Si metasurface decreases more rapidly than that of the Au metasurface, consistent with the numerical results shown in Figure 3 (a). Moreover, the suppression of the resonance for Si in water becomes more evident from figure 4(c). Additionally, the difference in the mean values of *Abs*, $(\overline{Abs_{Si}} - \overline{Abs_{Au}}) = \Delta\max(Abs)$ is plotted as a function of increasing $k_{env}$ (figure 4(d)). The experimentally calculated *Abs* difference closely follows the trend of the simulated results shown in Figure 3(d), with the *Abs* difference being positive in low-loss environments such as air, and negative in high-loss solvents. These observations indicate that the Si metasurface exhibits superior sensing performance under low-loss conditions, whereas the Au metasurface performs better in lossy environments.

## 3. Conclusion

Our results provide critical insights into the material choice for SEIRAS-based optofluidic sensing in realistic, lossy solvent environments. By employing BIC-driven metasurfaces to precisely control and decouple radiative damping from intrinsic material losses, we quantitatively demonstrate that neither silicon nor gold is universally superior in all solvent environments. To quantify this behavior experimentally, the metasurfaces were embedded in solvents such as DMSO, $D_2O$, and water and measured using a microfluidic setup. Our findings reveal a clear performance hierarchy: Si platforms excel in low-loss conditions (e.g., in air or low-loss environments with $k_{env} < 2.0 \times 10^{-3}$) while Au metasurfaces become the preferred choice in moderately lossy media. Crucially, we find that the transition point where both materials show matching performance is a function of the tilting angle, providing additional flexibility for sensor design. In highly lossy solvents like water ($k_{env} \approx 0.03$), both metasurfaces exhibit overall low resonance amplitudes and Q factors, limiting their sensitivity overall. Our observations provide a crucial roadmap for choosing the appropriate material platform for sensing in lossy environments, enabling the rational selection of metasurface materials tailored to specific solvent properties. Such solvent-optimized metasurface designs have the potential to significantly enhance the reliability and sensitivity of future SEIRA-based optofluidic systems for applications ranging from clinical diagnostics to fundamental biophysical studies in challenging liquid environments.

## 4. Materials and Methods

*Numerical Methods*

The simulations were performed using CST Studio Suite (Simulia), a commercial finite element solver. The setup included adaptive mesh refinement and periodic boundary conditions in the frequency domain. The refractive index of Si and CaF$_2$ were set to 3.449 and 1.4, respectively. Au was modelled according to the data provided by R. L. Olmon, B et. al.[45] and H$_2$O, D$_2$O according to the data provided by Jean-Joseph Max et. al. [46]. For the analyte, the refractive index was modelled as $n$ = 1.5 and $k$ = 0.36 according to the data provided by Zhang et. al[47] . The incident radiation was considered normal to the metasurface and with polarization parallel to the *x* axis for the simulations. The electric field enhancement was calculated using CST post-processing where the electric field at the middle of the resonators was numerically simulated and the average contribution of the electric field over the entire volume $P_x \times P_y \times (H+100\ nm)$ is evaluated.

*Fabrication*

For fabricating the Si metasurface, a 750 nm layer of amorphous silicon was deposited on a CaF$_2$ substrates using plasma-enhanced chemical vapour deposition with the PlasmaPro 100 system (Oxford Instruments). The nano-structuring process started with spin-coating a 400 nm layer of positive electron-beam resist, ZEP520A (Zeon Corporation), followed by a conductive polymer coating using ESPACER (Showa Denko K.K.). Electron-beam lithography was performed using an eLINE Plus system (Raith) at 20 kV with a 20 μm aperture. The patterned films were developed in an amyl acetate bath, followed by a bath of methyl isobutyl ketone and isopropyl alcohol (1:9 ratio). A 60 nm chromium layer was then deposited, and the resist was lifted off using Microposit Remover 1165 (Microresist). The remaining chromium served as an etching mask for the subsequent reactive ion etching process, which used SF6 and argon gases. Finally, the chromium mask was removed using TechniEtch Cr01 (MicroChemicals). The Au sample was fabricated by using the same nanostructuring process as the Si metasurface via electron beam lithography on CaF2. After patterning, an adhesion layer of 5 nm Ti and 100 nm of Au were deposited by electron-beam assisted evaporation, and the final metasurface structures were obtained by wet-chemical lift-off process (remover 1165, Microresist).

*Analyte Sensing*

The dual gradient metasurfaces were coated with 0.20% PMMA (495 K) diluted in anisole. The solution was uniformly applied to the metasurfaces through spin-coating at 3,000 rpm for 1 min. Following the coating, the metasurfaces were baked at 180 °C for 3 min to ensure the PMMA layer was fully solidified.

*Optical characterization*

Spectroscopic measurements were performed using a Spero microscope (Daylight Solutions Inc., USA) equipped with a 4×, 0.15 NA objective lens, providing a 2 × 2 mm$^2$ field of view. In reflectance mode, spectra were obtained in the range of 948 to 1800 cm$^{-1}$, with a spectral resolution of 2 cm$^{-1}$. Background measurements were performed on a gold mirror before the start of each measurement. The metasurface chip was mounted upside down in a home-made microfluidic cell, allowing the measurements through the backside of the substrate. The microfluidic cell used in this study comprises an inlet and an outlet. A syringe pump was used to control the flow of the sample solution inside the cell with a maximum flow rate of 500 μL min$^{-1}$. The pump was turned off during all spectroscopic measurements. The reflectance spectra were internally background-subtracted with ChemVision. In-house python code was then used to extract the metasurface spectra from the hyperspectral image data.

**Acknowledgements**

This project was funded by the Deutsche Forschungsgemeinschaft (DFG, German Research Foundation) under grant numbers EXC 2089/1–390776260 (Germany's Excellence Strategy) and TI 1063/1 (Emmy Noether Program), the Bavarian program Solar Energies Go Hybrid (SolTech), and Enabling Quantum Communication and Imaging Applications (EQAP), and the Center for NanoScience (CeNS). Funded by the European Union (ERC, METANEXT, 101078018 and EIC, NEHO, 101046329). Views and opinions expressed are however those of the author(s) only and do not necessarily reflect those of the European Union or the European Research Council Executive Agency. Neither the European Union nor the granting authority can be held responsible for them. S.A.M. additionally acknowledges the Lee-Lucas Chair in Physics.

**Data Availability Statement**

All data are available in the main text or the supplementary materials

## Author contributions

T.J., A.B., J.W. and A.T. conceived the idea and planned the research. T.J., A.B. and A.A. contributed to the sample fabrication. T.J., M.B., and A.B. performed the measurements. A.B., T.J. conducted the numerical simulations. T.J., A.B., M.B, and T.W. contributed to the data processing. A.A, T.J., M.B., contributed to the data analysis. S.A.M. and A.T. supervised the project. A.B. and T.J. wrote the manuscript with input from all the authors.

## Competing interests

The authors declare that they have no competing interests.

**Supporting Information**

All data are available in the main text or the supplementary materials.

**Supplementary information**

**A comparative analysis of plasmonic and dielectric metasurface sensing platforms powered by bound states in the continuum**

*Tao Jiang, Angana Bhattacharya, Martin Barkey, Andreas Aigner, Thomas Weber, Juan Wang, Stefan A. Maier, and Andreas Tittl\**

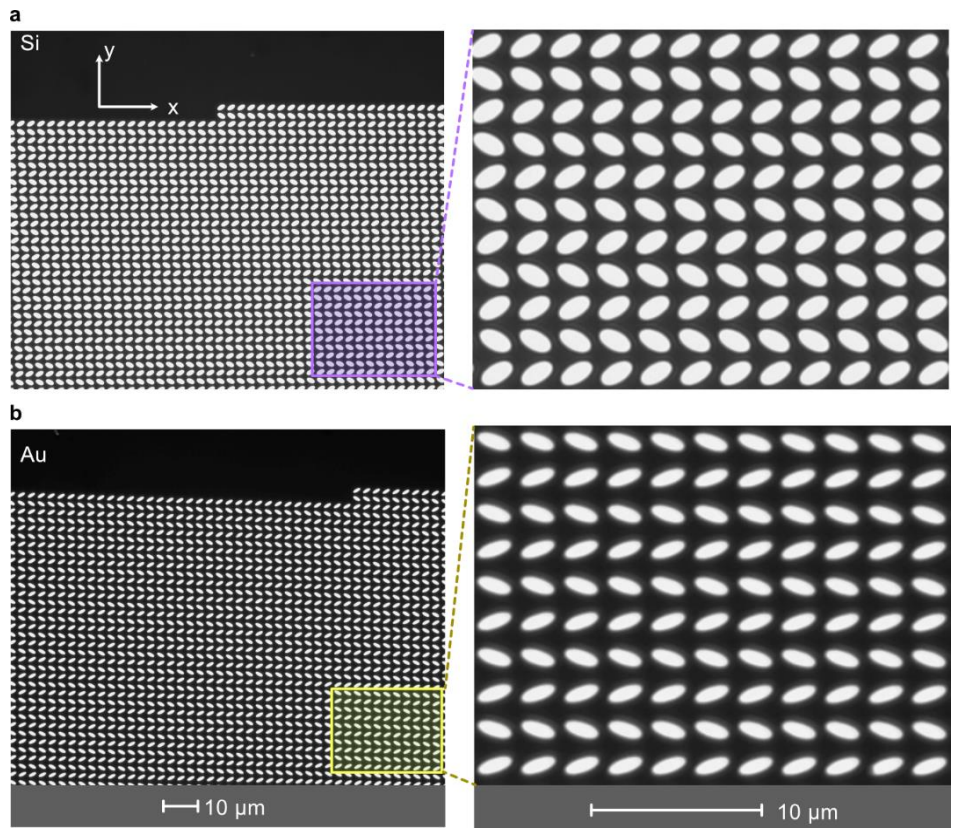

Figure S1. Optical microscopy images of (a) Si and (b) Au metasurfaces. The x direction corresponds to increasing structural scaling factors, while the y direction corresponds to varying tilting angles of the ellipses.

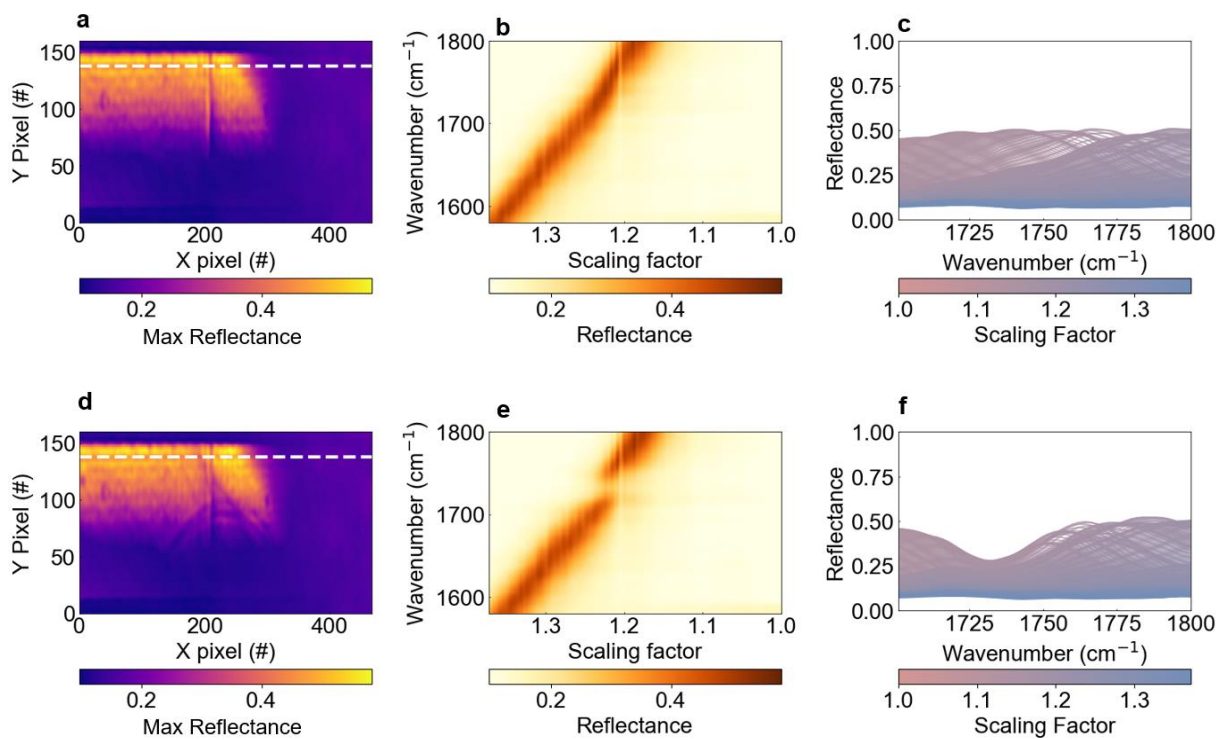

Figure S2. Reflectance spectra of the Au metasurface measured in air. Maximum reflectance spectra (a) without analyte and (d) with analyte. (b, e) Reflectance spectra along the white dashed lines in (a) and (d), respectively. Corresponding reflectance curves (c) without analyte and (f) with analyte. The dip at 1730 cm$^{-1}$ indicates the characteristic absorbance band of the analyte. The absorbance can be calculated from the reflectance (c) and (f) at 1730 cm$^{-1}$.

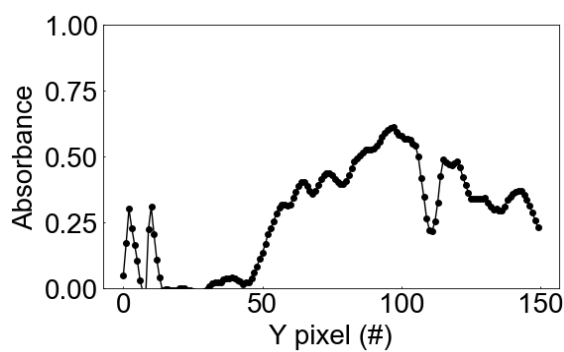

Figure S3. Absorbance of the Au metasurface in air at 1730 cm$^{-1}$ as a function of Y pixel position (corresponding to varying tilting angles). The tilting angle ranges from 0° to 30°. The maximum absorbance observed at pixel #91, corresponding to an angle of approximately 18°. The local dip in absorbance around pixel #100 is attributed to stitching effects from electron-beam lithography.

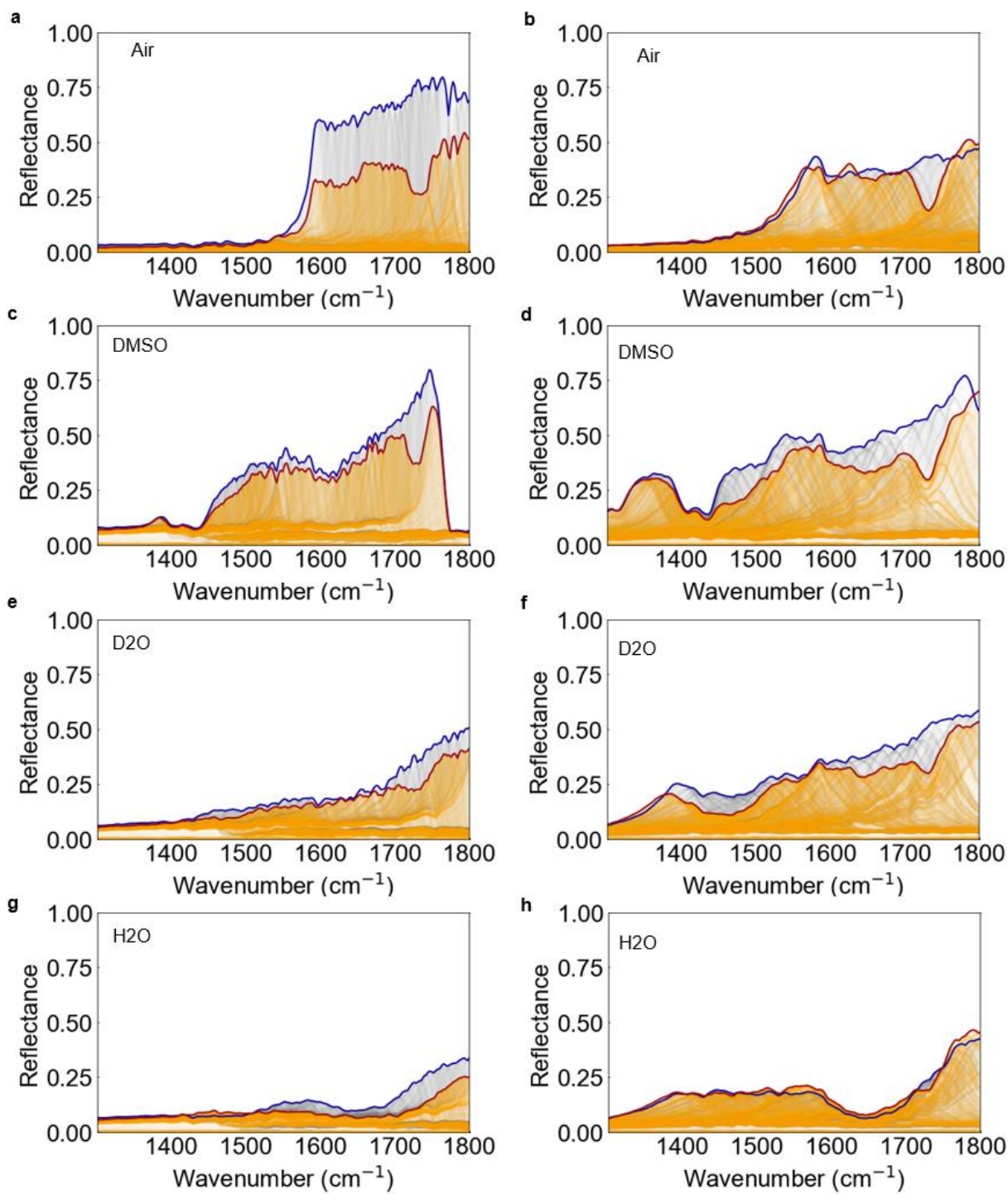

Figure S4. Full reflectance spectra of Si and Au metasurfaces without (gray curves with blue envelope) and with (orange curves with red envelope) analyte under various environmental conditions: (a, b) air, (c, d) dimethyl sulfoxide (DMSO), (e, f) deuterium oxide ($D_2O$), and (g, h) water ($H_2O$). The left column corresponds to the Si metasurface, and the right column corresponds to the Au metasurface. As the absorption of the solvent environment increases, progressing from DMSO to $D_2O$ to $H_2O$, the reflectance spectra exhibit a corresponding reduction in intensity.